\documentclass[aps,preprint]{revtex4}%
\usepackage{amsfonts}
\usepackage{amsmath}
\usepackage{amssymb}
\usepackage{graphicx}%
\begin{document}
\preprint{ }
\title{Mass inequality for the quark propagator}
\author{Dean Lee and Richard Thomson}
\affiliation{Department of Physics, North Carolina State University, Raleigh, NC 27695-8202}

\begin{abstract}
We show that for any gauge-fixing scheme with positive semi-definite
functional integral measure, the inverse correlation length \noindent of the
quark propagator is bounded below by one-half the pion mass.

\end{abstract}
\date{\today}
\maketitle

It is known that quarks develop a dynamically-generated mass in the Nambu
Jona-Lasinio model, a self-interacting relativistic quark model without quark
confinement \cite{Nambu:1961tp,Nambu:1961fr,Vogl:1989ea}. \ Studies using the
Dyson-Schwinger or gap equation for the quark propagator suggest a similar
phenomenon occurs in quantum chromodynamics (QCD)
\cite{Roberts:1994dr,Alkofer:2000wg,Maris:2003vk,Bhagwat:2003vw}. \ For even
light quarks one finds a rather large inverse correlation length or
nonperturbative mass function, $M$, at zero momentum.

This dressed quark mass could be related to the constituent quark mass of the
na\"{\i}ve nonrelativistic quark model. \ Dyson-Schwinger studies suggest a
value for $M(0)$ somewhere between $170$ MeV and $500$ MeV. \ Recent quenched
lattice results using Landau and Laplacian gauge with the overlap fermions
\cite{Zhang:2004gv} and improved staggered fermions \cite{Bowman:2002bm} and
unquenched lattice results using Landau gauge with improved staggered fermions
\cite{Bowman:2005vx} yield a similar value for $M(0)$ between $250$ and $500$ MeV.

In this work we introduce another tool for understanding the infrared behavior
of quark propagators in QCD. \ We prove a general inequality which relates
$M(0)$ and the pion mass, $m_{\pi}$. \ The inequality applies to all
gauge-fixing schemes with positive semi-definite functional integral measure.

Weingarten showed that for an even number of degenerate quark flavors, the
pion must have the lowest energy of all meson states \cite{Weingarten:1983uj}.
\ The proof required positivity of the QCD action, $\gamma_{5}$-Hermitivity,
and exact quark flavor symmetry. \ With exactly the same conditions for a
gauge-fixed action, we show that%
\begin{equation}
\left[  M(0)\right]  _{\text{gauge-fixed}}\geq\frac{1}{2}\left[  m_{\pi
}\right]  _{\text{gauge-fixed}}.
\end{equation}
The requirement of positivity is satisfied by all gauge-fixing schemes used in
recent lattice simulations. \ See \cite{Giusti:2001xf,Williams:2003du} for
reviews of current lattice gauge-fixing methods.

By positivity we mean positivity of the Euclidean functional integral measure
and not spectral positivity in a Hamiltonian or transfer matrix formalism.
\ While spectral positivity is violated in certain gauges
\cite{Philipsen:2001hd,Aubin:2004av}, any gauge-fixing scheme without a
positive functional integral measure would have a sign or complex phase
problem and be of limited practical use. \ In the following we say nothing new
about the subject of incomplete gauge fixing and Gribov copies for any
particular scheme \cite{Gribov:1977wm,vanBaal:1997gu}. \ But if the Gribov
problem is in fact handled properly for a given gauge-fixing scheme then we
can conclude further that%
\begin{equation}
\left[  M(0)\right]  _{\text{gauge-fixed}}\geq\frac{1}{2}\left[  m_{\pi
}\right]  _{\text{gauge-fixed}}=\frac{1}{2}m_{\pi}.
\end{equation}

We begin by establishing our notation for QCD with $N_{c}$ colors. \ We let
$\lambda^{A}$ be the matrix generators for $SU(N_{c})$, and we use the matrix
field notation,%
\begin{align}
\mathbf{A}_{\mu}  &  =A_{\mu}^{B}\frac{\lambda^{B}}{2},\\
\mathbf{F}_{\mu\nu}  &  =F_{\mu\nu}^{B}\frac{\lambda^{B}}{2},
\end{align}
where
\begin{equation}
F_{\mu\nu}^{B}=\partial_{\mu}A_{\nu}^{B}-\partial_{\nu}A_{\mu}^{B}%
-g_{0}f^{BCD}A_{\mu}^{C}A_{\nu}^{D}.
\end{equation}
Summations are implied by repeated indices. \ We consider $n_{f}$ degenerate
quark flavors where $n_{f}$ is even, and label two of the flavors as $u$ and
$d.$ \ In Euclidean space, the QCD action prior to gauge fixing is%
\begin{equation}
S=S_{g}+\sum_{i=1}^{n_{f}}\int d^{4}x\;\bar{q}_{i}\left(  \gamma_{\mu}%
^{E}(\partial_{\mu}+ig_{0}\mathbf{A}_{\mu})+m_{0}\right)  q_{i},
\end{equation}
where%
\begin{equation}
S_{g}=\frac{1}{2}Tr\left[  \int d^{4}x\;\mathbf{F}_{\mu\nu}\mathbf{F}_{\mu\nu
}\right]  .
\end{equation}
Throughout we assume a gauge-invariant lattice discretization, though we use
the more familiar continuum notation.

Let $M$ be the quark matrix due to the background gluon field. \ $M$ has the
block-diagonal structure%
\begin{equation}
M=%
%TCIMACRO{\dbigoplus \limits_{i=1}^{n_{f}}}%
%BeginExpansion
{\displaystyle\bigoplus\limits_{i=1}^{n_{f}}}
%EndExpansion
M^{q_{i}},
\end{equation}
where all blocks $M^{q_{i}}$ are the same,%
\begin{equation}
M^{q_{i}}=M^{q}\equiv\gamma_{\mu}^{E}(\partial_{\mu}+ig_{0}\mathbf{A}_{\mu
})+m_{0}.
\end{equation}
We consider any lattice quark action so long as $\gamma^{5}$-Hermitivity is
maintained and the quark flavor symmetry is exact. \ For most quark actions
this holds true at any lattice spacing. \ For staggered quarks, however, one
must go to the continuum limit in order to regain exact flavor symmetry.

The condition of $\gamma^{5}$-Hermitivity means%
\begin{equation}
\gamma^{5}(M^{q})^{\dagger}\gamma^{5}=M^{q}\text{,}%
\end{equation}
and so $\det M^{q}$ must be real.\ \ The block-diagonal structure and flavor
symmetry then gives%
\begin{equation}
\det M=\left(  \det M^{q}\right)  ^{n_{f}}\geq0\text{.}%
\end{equation}
$M^{q}$ has several types of indices corresponding with space-time lattice
points, Dirac indices, and color indices in the fundamental representation.
\ When needed we use the notation%
\begin{equation}
M_{ij;\mathbf{ab}}^{q}(x,y)
\end{equation}
where $x,y$ are the space-time lattice points; $i,j$ are the Dirac indices;
and $\mathbf{a},\mathbf{b}$ are the color indices.

We fix the gauge using the constraint functional $f(A)$ and the
gauge-invariant Faddeev-Popov determinant $\Delta_{f}(A)$
\cite{Giusti:2001xf,Williams:2003du}. \ $\Delta_{f}(A)$ is defined by the
condition%
\begin{equation}
\Delta_{f}^{-1}(A)=\int DG\,\delta(f(A^{G})),
\end{equation}
where the integration is over all gauge transformations and $A^{G}$ is the
gauge orbit of $A$. \ We can write $\Delta_{f}^{-1}(A)$ as a sum over all
solutions $G_{i}$ to the constraint equation $f(A^{G})=0,$%
\begin{equation}
\Delta_{f}^{-1}(A)=\sum_{i}\frac{1}{\left\vert \det\frac{\delta f(A^{G}%
)}{\delta G}\right\vert _{G=G_{i}}}. \label{determinant}%
\end{equation}
$\Delta_{f}(A)$ defined in this way is clearly positive semi-definite. \ If
$f$ is an ideal gauge-fixing constraint, then there is exactly one simple zero
$f(A^{G})=0$ for each gauge orbit. Otherwise we must contend with possible
complications such as singular configurations where one or more of the
determinants in (\ref{determinant}) vanish or perhaps even the number of
Gribov copies becomes infinite.

We note that
\begin{align}
\Delta_{f}(A^{G})  &  =\Delta_{f}(A),\\
S_{g}(A^{G})  &  =S_{g}(A),\\
\det[M(A^{G})]  &  =\det[M(A)],
\end{align}
and for any gauge-invariant observable,%
\begin{equation}
O(A^{G})=O(A).
\end{equation}
Therefore the expectation value of any gauge-invariant observable is%
\begin{align}
\left\langle O\right\rangle  &  =\frac{\int DA\;O(A)e^{-S_{g}(A)}\det
[M(A)]}{\int DA\;e^{-S_{g}(A)}\det[M(A)]}\nonumber\\
&  =\frac{\int DA\;O(A)e^{-S_{g}(A)}\det[M(A)]\Delta_{f}(A)\int DG\,\delta
(f(A^{G}))}{\int DA\;e^{-S_{g}(A)}\det[M(A)]\Delta_{f}(A)\int DG\,\delta
(f(A^{G}))}\nonumber\\
&  =\frac{\int DA\;O(A)e^{-S_{g}(A)}\det[M(A)]\Delta_{f}(A)\,\delta
(f(A))}{\int DA\;e^{-S_{g}(A)}\det[M(A)]\Delta_{f}(A)\,\delta(f(A))}.
\label{expect}%
\end{align}
For observables which are not gauge-invariant, we use the last expression in
(\ref{expect})\ to define the gauge-fixed expectation value,%
\begin{equation}
\left\langle O\right\rangle _{f}\equiv\frac{\int DA\;O(A)e^{-S_{g}(A)}%
\det[M(A)]\Delta_{f}(A)\,\delta(f(A))}{\int DA\;e^{-S_{g}(A)}\det
[M(A)]\Delta_{f}(A)\,\delta(f(A))}. \label{gauge fixed}%
\end{equation}
We note that this gauge-fixed functional integral measure is positive semi-definite.

Unfortunately the action in (\ref{gauge fixed}) is too difficult to deploy in
actual lattice simulations. \ In practice, gauge fixing on the lattice is
implemented using
\begin{equation}
\left\langle O\right\rangle _{f}\equiv\frac{\int DA\;O(A^{G(A)})e^{-S_{g}%
(A)}\det[M(A)]}{\int DA\;e^{-S_{g}(A)}\det[M(A)]}%
\end{equation}
where $G(A)$ is a special gauge transformation for which $f(A^{G(A)})=0.$
\ This gauge-fixed functional integral measure is also positive semi-definite.
\ If $f$ is an ideal gauge-fixing constraint, then there is exactly one
solution $f(A^{G})=0$ for each gauge orbit and the process of selecting $G(A)$
is unique. \ Otherwise we must again contend with complications due to
multiple Gribov copies.

Following Weingarten \cite{Weingarten:1983uj}\ we consider the pion
correlation function%
\begin{equation}
F_{\pi}(x)=\left\langle \bar{d}_{i\mathbf{a}}(x)\gamma_{ij}^{5}u_{j\mathbf{a}%
}(x)\;\bar{u}_{k\mathbf{b}}(0)\gamma_{kl}^{5}d_{l\mathbf{b}}(0)\right\rangle .
\end{equation}
We can write%
\begin{equation}
F_{\pi}(x)=\frac{\int DADqD\bar{q}\;\bar{d}_{i\mathbf{a}}(x)\gamma_{ij}%
^{5}u_{j\mathbf{a}}(x)\;\bar{u}_{k\mathbf{b}}(0)\gamma_{kl}^{5}d_{l\mathbf{b}%
}(0)e^{-S_{g}(A)}\Delta_{f}(A)\,\delta(f(A))}{\int DADqD\bar{q}\;e^{-S_{g}%
(A)}\Delta_{f}(A)\,\delta(f(A))}.
\end{equation}
Integrating over the quark fields we get
\begin{equation}
F_{\pi}(x)=-\int D\Theta\;(M^{q})_{li;\mathbf{ba}}^{-1}(0,x)\gamma_{ij}%
^{5}(M^{q})_{jk;\mathbf{ab}}^{-1}(x,0)\gamma_{kl}^{5},
\end{equation}
where $D\Theta$ is the probability measure,%
\begin{equation}
D\Theta=\frac{DAe^{-S_{G}(A)}\det[M(A)]\Delta_{f}(A)\,\delta(f(A))}{\int
DAe^{-S_{G}(A)}\det[M(A)]\Delta_{f}(A)\,\delta(f(A))}.
\end{equation}
Since%
\begin{equation}
\gamma^{5}(M^{q})^{-1}\gamma^{5}=((M^{q})^{-1})^{\dagger}\text{,}%
\end{equation}
we have%
\begin{equation}
F_{\pi}(x)=-\sum_{i,l,\mathbf{a},\mathbf{b}}\int D\Theta\;\left\vert
(M^{q})_{li;\mathbf{ba}}^{-1}(0,x)\right\vert ^{2}.
\end{equation}

Next we consider the quark correlator,%
\begin{equation}
F_{q}(x)=\left\langle \bar{u}_{k\mathbf{a}}(x)\Gamma_{kj}u_{j\mathbf{a}%
}(0)\right\rangle ,
\end{equation}
where $\Gamma$ is an arbitrary matrix in Dirac spinor space. \ We can write%
\begin{equation}
F_{q}(x)=\frac{\int DADqD\bar{q}\;\bar{u}_{k\mathbf{a}}(x)\Gamma
_{kj}u_{j\mathbf{a}}(0)e^{-S_{g}(A)}\Delta_{f}(A)\,\delta(f(A))}{\int
DADqD\bar{q}\;e^{-S_{g}(A)}\Delta_{f}(A)\,\delta(f(A))}.
\end{equation}
Integrating over the quark fields we get%
\begin{align}
F_{q}(x)  &  =-\int D\Theta\;(M^{q})_{jk;\mathbf{aa}}^{-1}(0,x)\Gamma
_{kj}\nonumber\\
&  =-N_{c}\int D\Theta\;(M^{q})_{jk;\mathbf{aa}}^{-1}(0,x)\Gamma_{kj}\text{
(no sum on }\mathbf{a}\text{).}%
\end{align}
Let us define
\begin{equation}
\gamma=\sqrt{\sum\limits_{k,j}\left\vert \Gamma_{kj}\right\vert ^{2}}.
\label{gamma}%
\end{equation}
By the Cauchy-Schwarz inequality we have%
\begin{align}
\left\vert F_{q}(x)\right\vert  &  \leq N_{c}\gamma\sqrt{\sum_{j,k}\int
D\Theta\;\left\vert (M^{u})_{jk;\mathbf{aa}}^{-1}(0,x)\right\vert ^{2}}\text{
(no sum on }\mathbf{a}\text{)}\nonumber\\
&  \leq N_{c}\gamma\sqrt{\frac{1}{N_{c}}\sum_{j,k,\mathbf{a}}\int
D\Theta\;\left\vert (M^{u})_{jk;\mathbf{aa}}^{-1}(0,x)\right\vert ^{2}}.
\end{align}
So therefore%
\begin{equation}
\left\vert F_{q}(x)\right\vert \leq\gamma\sqrt{N_{c}}\sqrt{\left\vert F_{\pi
}(x)\right\vert }. \label{corrineq}%
\end{equation}

We now take $x_{1}=x_{2}=x_{3}=0$ and consider the limit as $x_{4}%
\rightarrow\pm\infty$. \ For any correlation function $F(x)$ we define an
inverse correlation length as the supremum of the set of real numbers $\alpha$
such that%
\begin{equation}
\lim_{x_{4}\rightarrow+\infty}\left\vert F(x)\right\vert e^{\alpha\left\vert
x_{4}\right\vert }=\lim_{x_{4}\rightarrow-\infty}\left\vert F(x)\right\vert
e^{\alpha\left\vert x_{4}\right\vert }=0\text{.}%
\end{equation}
We are implicitly assuming that the set of $\alpha$'s is not empty and bounded
above. \ While this seems a safe assumption, it is perhaps not completely
trivial given the lack of spectral positivity.

We define $\left[  m_{\pi}\right]  _{f}$ as the inverse correlation length of
$F_{\pi}(x)$ and $\left[  M(0)\right]  _{f}$ as the inverse correlation length
of $F_{q}(x)$. \ We conclude from (\ref{corrineq})\ that%
\begin{equation}
\left[  M(0)\right]  _{f}\geq\frac{1}{2}\left[  m_{\pi}\right]  _{f}\text{.}%
\end{equation}
If we assume that our gauge-fixing scheme does not affect the calculation of
the gauge-invariant pion correlator, then we also have
\begin{equation}
\left[  M(0)\right]  _{f}\geq\frac{1}{2}\left[  m_{\pi}\right]  _{f}=\frac
{1}{2}m_{\pi}. \label{massineq}%
\end{equation}

The mass inequality in (\ref{massineq}) should hold for both quenched QCD and
unquenched QCD. \ It holds true at nonzero lattice spacing, any quark mass,
any number of colors, any even number of flavors, and any gauge-fixing scheme
so long as we have a positive semi-definite functional integral measure,
$\gamma^{5}$-Hermitivity, and exact flavor symmetry. \ In the chiral limit,
the inequality does not require that $\left[  M(0)\right]  _{f}$ be nonzero.
\ Nor does it imply that $\left[  M(0)\right]  _{f}$ is the same for different
gauge-fixing schemes. \ However it does require that $\left[  M(0)\right]
_{f}$ be at least as large as $\frac{1}{2}m_{\pi}$, which is proportional to
the square root of the current quark mass in unquenched QCD and a slightly
modified fractional power in quenched QCD due to quenched chiral logarithms
\cite{Sharpe:1992ft,Bernard:1992mk}.

If QCD did not confine quarks, then our mass inequality would be rather
trivial. \ It would simply say that if the pion exists, then its mass must lie
below the quark-antiquark continuum threshold. \ However in the real world the
quark propagator is not directly observable and quarks are confined.
\ Therefore the mass inequality is no longer trivial. \ If, in some
gauge-fixing scheme, $\left[  M(0)\right]  _{f}$ could be identified with a
constituent quark mass, then (\ref{massineq}) implies that the pion mass lies
below the constituent quark-antiquark threshold.

Recent results from the Dyson-Schwinger equation and lattice simulations at
various quark masses suggest that the mass inequality in (\ref{massineq}) is
indeed satisfied. \ Further calculations would be useful to understand the
dependence of $\left[  M(0)\right]  _{f}$ on gauge-fixing scheme, lattice
spacing, quark masses, different quark actions, number of colors and flavors,
and quenching artifacts.

\textit{The authors thank Steve Cotanch, Pieter Maris, Craig Roberts, Thomas
Schaefer, Pierre van Baal, and Tony Williams for reading earlier versions of
the manuscript and for helpful discussions and comments. \ This work was
supported by Department of Energy grant DE-FG02-04ER41335.}

\bibliographystyle{apsrev}
\bibliography{NuclearMatter}

\end{document}